\documentclass[a4paper,11pt]{article}
\usepackage{pos}
\usepackage{subcaption}
\title{TES Detector for ALPS II}

\author*[a]{Rikhav Shah}
\author[b]{Katharina-Sophie Isleif}
\author[b]{Friederike Januschek}
\author[b]{Axel Lindner}
\author[a]{Matthias Schott}

\affiliation[a]{Johannes Gutenberg-Universität Mainz,\\
	Staudingerweg 7, 55128 Mainz, Germany}

\affiliation[b]{Deutsches Elektronen-Synchrotron DESY,\\
	Notkestr. 85, 22607 Hamburg, Germany}

\emailAdd{rshah@uni-mainz.de}

\abstract{The application of cryogenic single photon detectors has found great use in high precision particle  physics experiments such as ALPS (Any Light Particle Search) II, which implements it for fundamental studies to search for new particles. ALPS II is a light-shining-through-a-wall experiment searching for axion-like-particles, which couple to photons. The extremely low rate of photons generated by the conversion of such axion-like-particles necessitates a detector setup capable of low energy ($\sim$1\,eV; as dictated by cavity optics) single photon detection with high efficiency and an ultra low background level, with long-term stability. This can be realised by a Transition Edge Sensor (TES) setup with low temperature SQUID readout.}

\FullConference{%
	*** The European Physical Society Conference on High Energy Physics (EPS-HEP2021), ***\\
	*** 26-30 July 2021 ***\\
	*** Online conference, jointly organized by Universität Hamburg and the research center DESY ***
}

\begin{document}
	\maketitle
	
	\section{Introduction}
	
	Axions are pseudoscalar Nambu-Goldstone bosons, first hypothesised to remedy the CP problem in QCD (established in \cite{ref_1}-\cite{ref_3}). Along with their cousins, the axion-like-particles (unrelated to the strong CP problem), they could contribute substantially also to dark matter \cite{ref_4}, while axion-like-particles are also hinted at by other astrophysical phenomena \cite{ref_5}. ALPS II at DESY, Hamburg, is the successor to the ALPS I experiment \cite{ref_7}, and uses a light-shining-through-a-wall concept to generate and detect axions \cite{ref_14} and axion-like-particles. The ALPS II experiment consists of laser light (1064\,nm wavelength) built up to high power in an optical cavity of length $\sim$120\,m with magnetic field of 5.3\,T, where axion-like-particles may be produced via the Primakoff-like Sikivie effect \cite{ref_6}. After crossing a light-tight barrier or wall, these can reconvert to photons via the same mechanism in an optical cavity (resonating to the same mode as the former cavity) and subsequently be detected. The rate of production of such single photons from axion/axion-like-particle decay is extremely low, about $10^{-5}$\,cps (approx. 1-2 photons per day) \cite{ref_8} while striving for an axion-photon coupling $g_{a \gamma \gamma} \approx 2 \cdot 10^{-11} $\,GeV$^{-1}$ with ALPS II. Combined with the low energy of the detected photon (1.165\,eV), the ultra low background level required ( $<10^{-5}$\,cps), and the high detection efficiency needed, the detection of these photons is a unique challenge. We have demonstrated that the intrinsic background rate of the Transition Edge Sensor (TES) can meet this goal.
	
	\section{TES Detector and Setup}
	
\begin{figure}
		\centering
		\begin{subfigure}[t]{0.4\textwidth}
			\includegraphics[width=\textwidth]{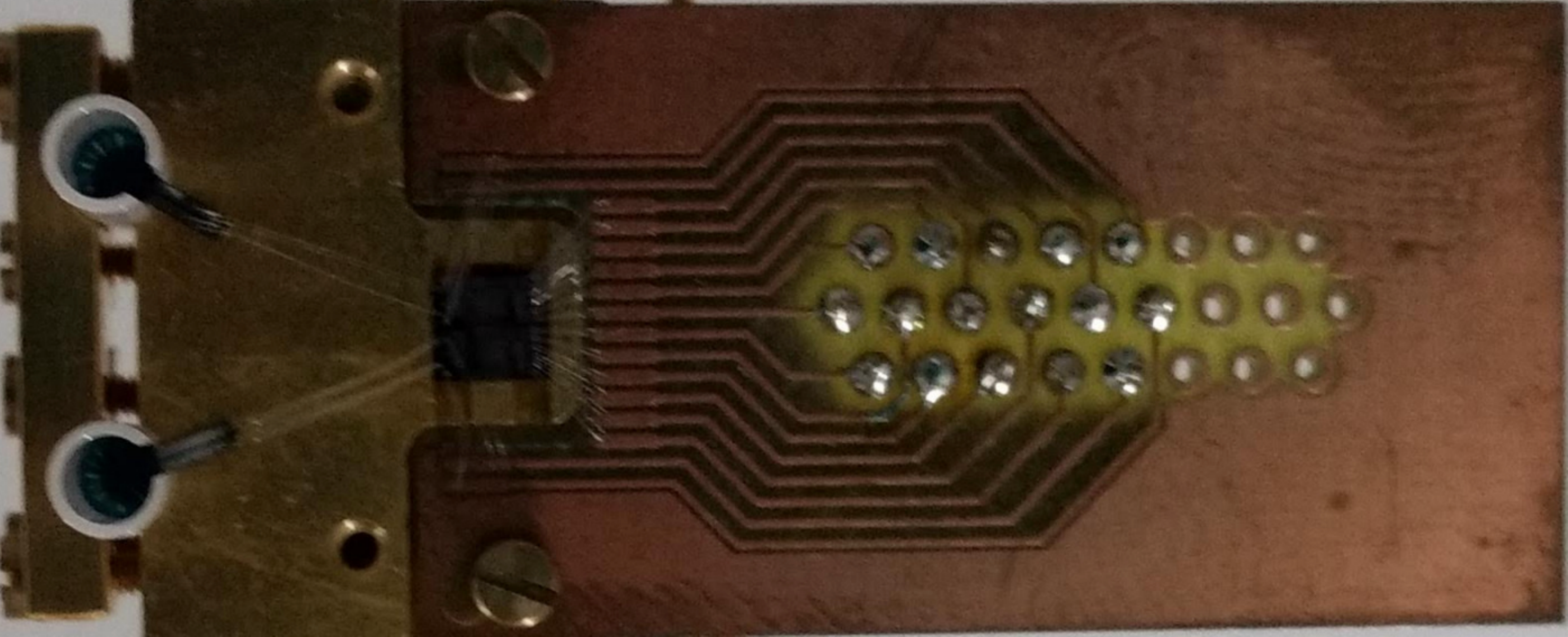}
			\caption{The TES detector module from PTB with the two TESs (inside white ferrules, fabricated by NIST) and their bond wires to the SQUIDs, also integrated on the module by PTB.\\
				Approx. size of module: 1\,cm x 2\,cm.}
			\label{Figure: TESModule}
		\end{subfigure}
		\hfill
		\begin{subfigure}[t]{0.5\textwidth}
			\includegraphics[width=\textwidth]{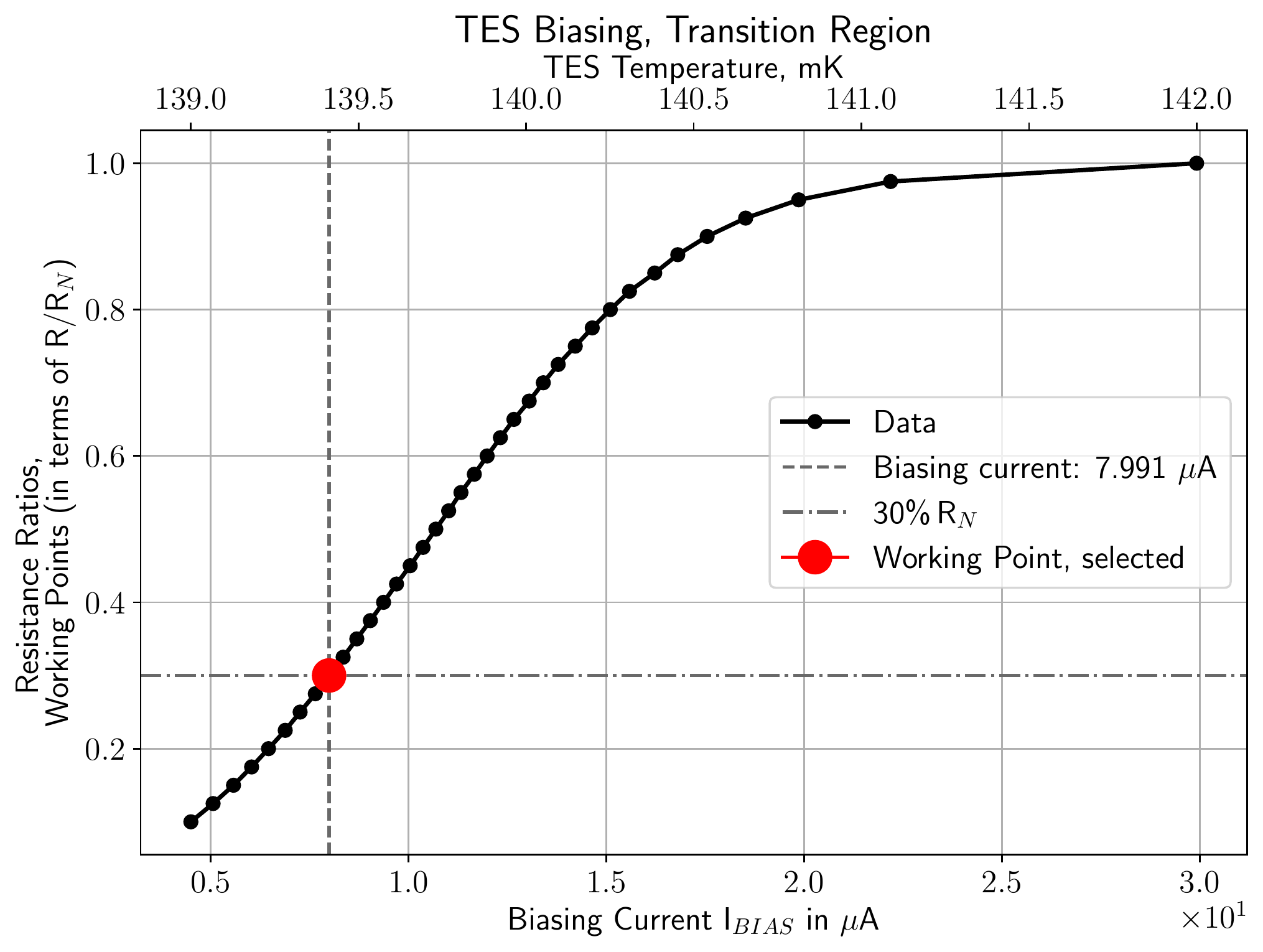}
			\caption{The TES is biased to operate it in its transition region. The selected working point corresponds to 30\% of the normal conducting resistance $R_N$, and the working points are described as a ratio of this resistance $R_N$. The change of the TES resistance $R_{TES}$ with its temperature is also shown.}
			\label{Figure: TESBiasing}
		\end{subfigure}
		\caption{}
	\end{figure}

	A TES is a cryogenic superconducting microcalorimeter operating in its transition region \cite{ref_10}, where the drastic dependence of the TES resistance on the temperature is used. We use a TES microchip from NIST (National Institute of Standards and Technology, USA) operated around its critical temperature ($T_C$) of 140\,mK by biasing it appropriately, as in Figure \ref{Figure: TESBiasing}. A fiber coupled setup can reach detection efficiencies of 95\% \cite{ref_9}. The absorption of a single 1064\,nm photon increases the temperature by $\sim$320\,$\mu$K and the resistance by $\sim$7\,$\Omega$  at the selected working point (following also \cite{ref_12}). 
	The consequent change in the current can be measured by SQUIDs (Superconducting Quantum Interference Devices, acting as sensitive magnetometers) which are integrated on the detector module by PTB (Physikalisch-Technische Bundesanstalt), Germany (Figure \ref{Figure: TESModule}). These are operated and read out by SQUID electronics from Magnicon GmbH. The entire setup is housed in a dilution refrigerator from BlueFors cooling it to a base temperature $<25$\,mK. With this setup, we can characterise signals and backgrounds for ALPS II.
	
	\section{TES Characterisation}
	
\begin{figure}[h]
		\centering
		\begin{subfigure}[t]{0.48\textwidth}
			\includegraphics[width=\textwidth]{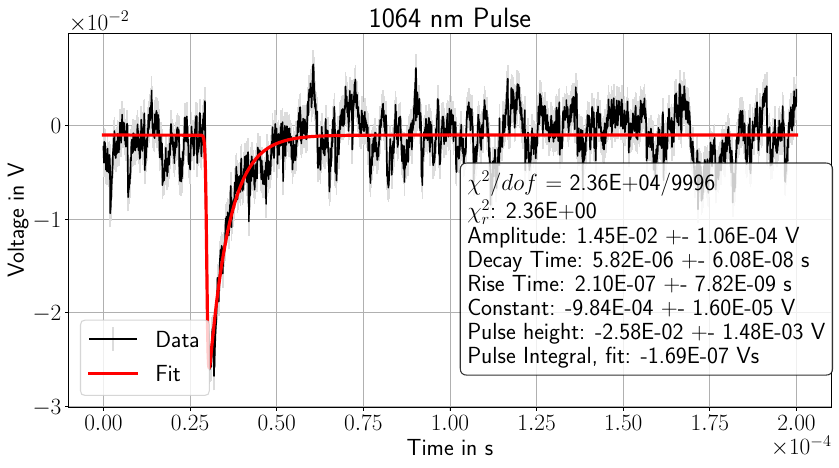}
			\caption{A 1064\,nm photon pulse as triggered in the TES and read out by the SQUID electronics and further DAQ system. The values of the fit parameters and pulse characteristics are also given.}
			\label{Figure:TESLight}
		\end{subfigure}
		\hfill
		\centering
		\begin{subfigure}[t]{0.48\textwidth}
			\includegraphics[width=\textwidth]{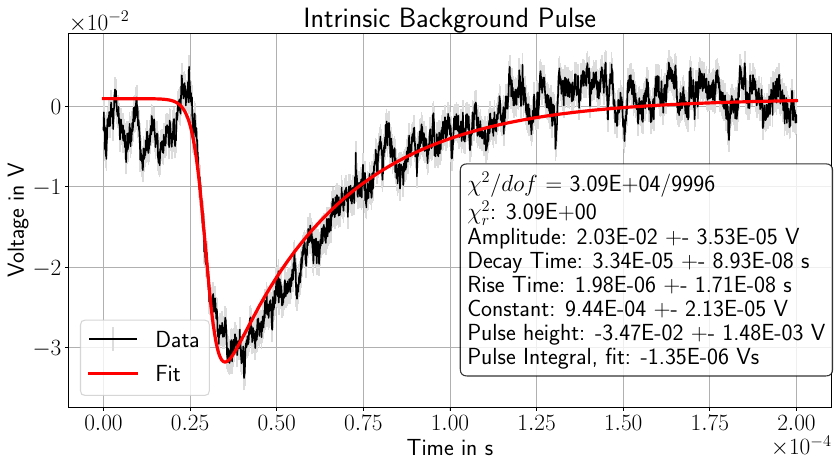}
			\caption{An intrinsic background pulse as triggered in the TES without an optical fiber attached to it. The shown values of the fit parameters describe the difference from a 1064\,nm pulse, in this single example.}
			\label{Figure:TESInt}
		\end{subfigure}
		\caption{}
	\end{figure}

	The input of 1064\,nm photons from a continuous wave (cw) laser to the TES via the optical fiber is used to calibrate the response of the biased TES. These events, triggered in the detector (shown in Figure \ref{Figure:TESLight}), can be fitted with a modified TES response function, the exemplary values of which are also shown in Figure \ref{Figure:TESLight}. 
	Intrinsic background pulses can be triggered in the TES (Figure \ref{Figure:TESInt}) with a dark system with no fiber connected. The backgrounds could be due to radioactivity, cosmic rays, photons from black-body-radiation (expected to dominate the background seen with an optical fiber attached to the TES), electromagnetic fields, etc. Other sources like Cherenkov radiation, transition radiation, etc., (which could be caused by cosmic rays as well) \cite{ref_17} will also be investigated. The fit parameters of these backgrounds (following a purely phenomenological perspective) may vary significantly from those of 1064\,nm light pulses, allowing us to reject such background events.
	
	\section{Pulse Discrimination}
\begin{figure}
		\centering
		\begin{subfigure}[t]{0.48\textwidth}
			\includegraphics[width=\textwidth]{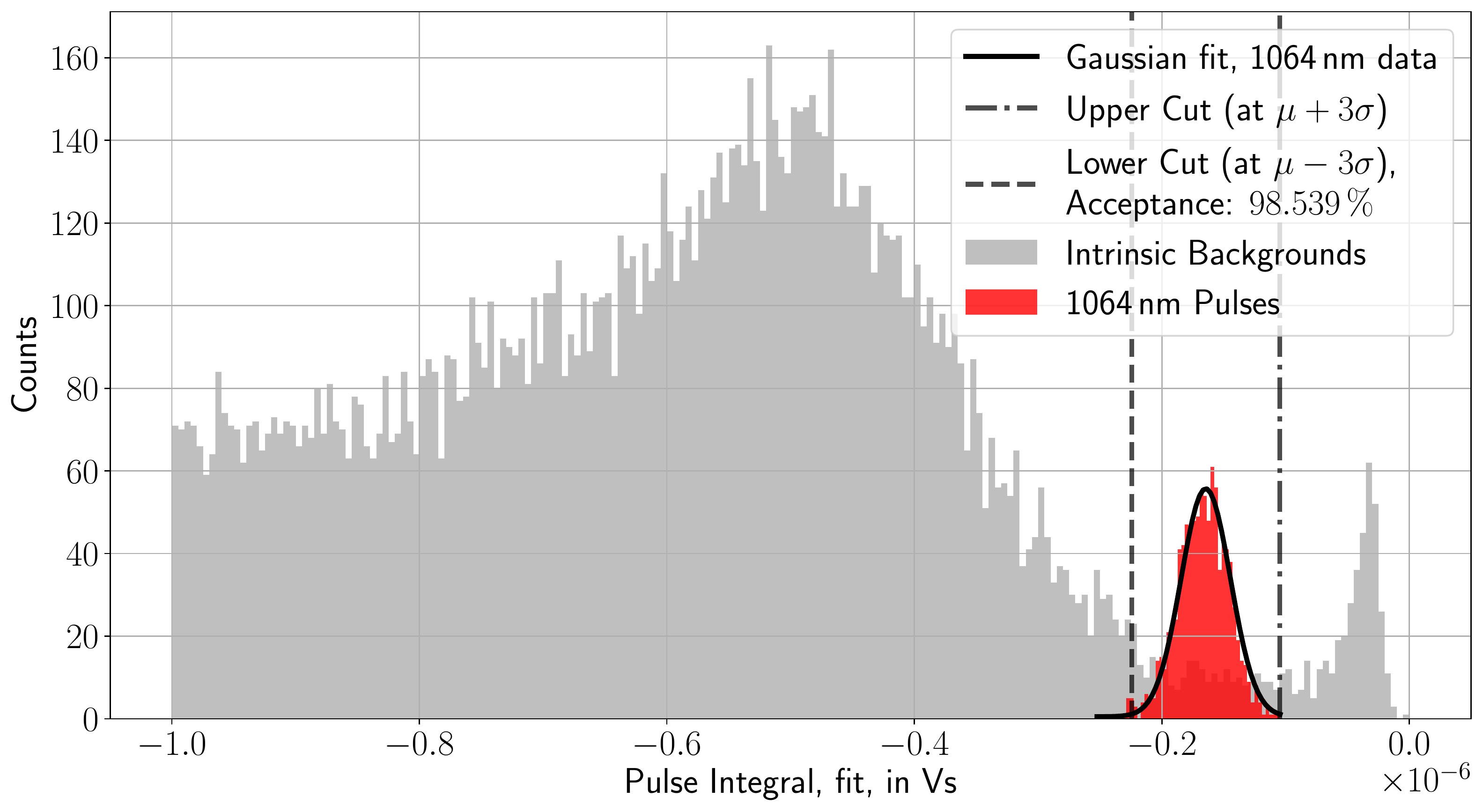}
			\caption{}
		\end{subfigure}
		\begin{subfigure}[t]{0.48\textwidth}
			\includegraphics[width=\textwidth]{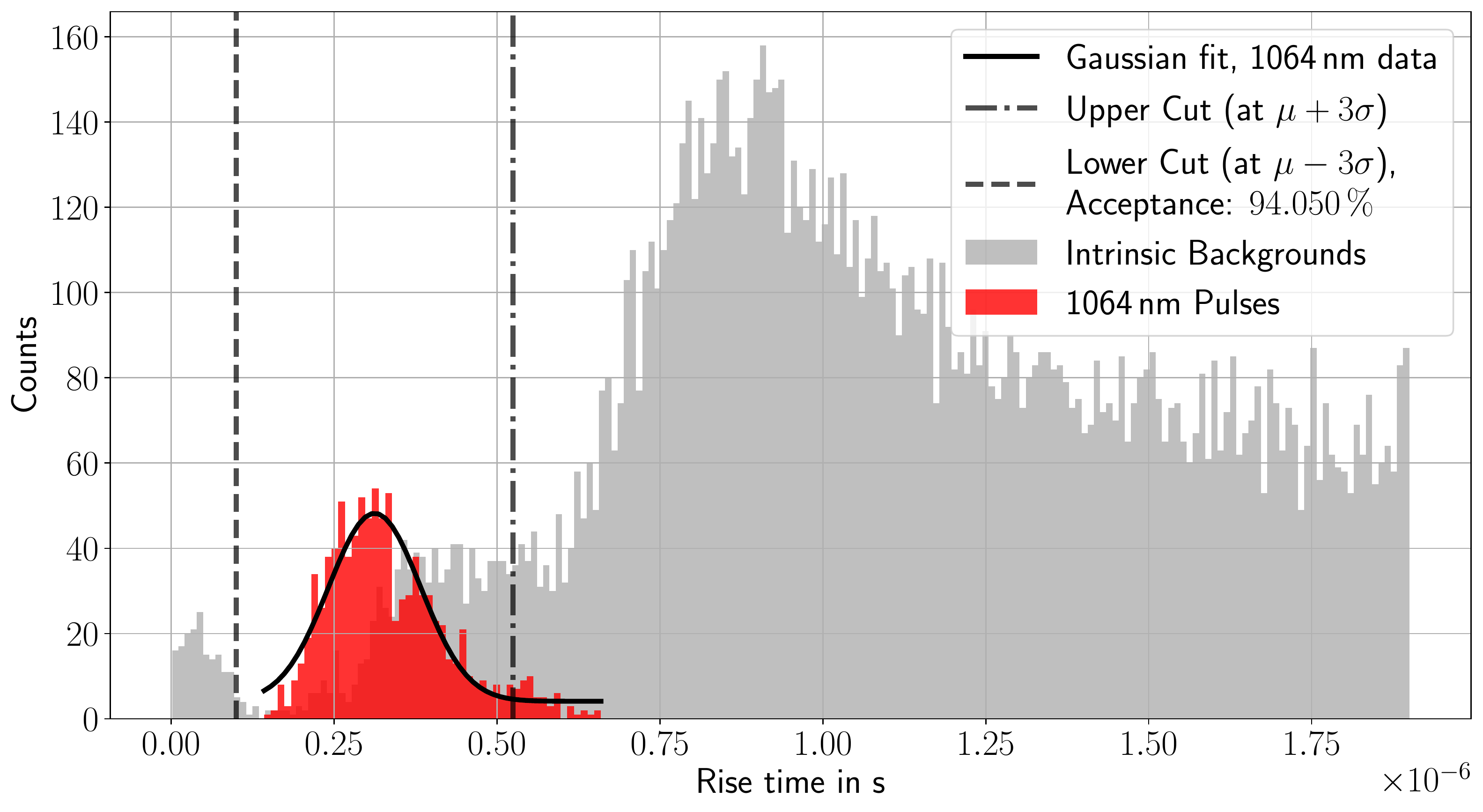}
			\caption{}
		\end{subfigure}
		\begin{subfigure}[t]{0.48\textwidth}
			\includegraphics[width=\textwidth]{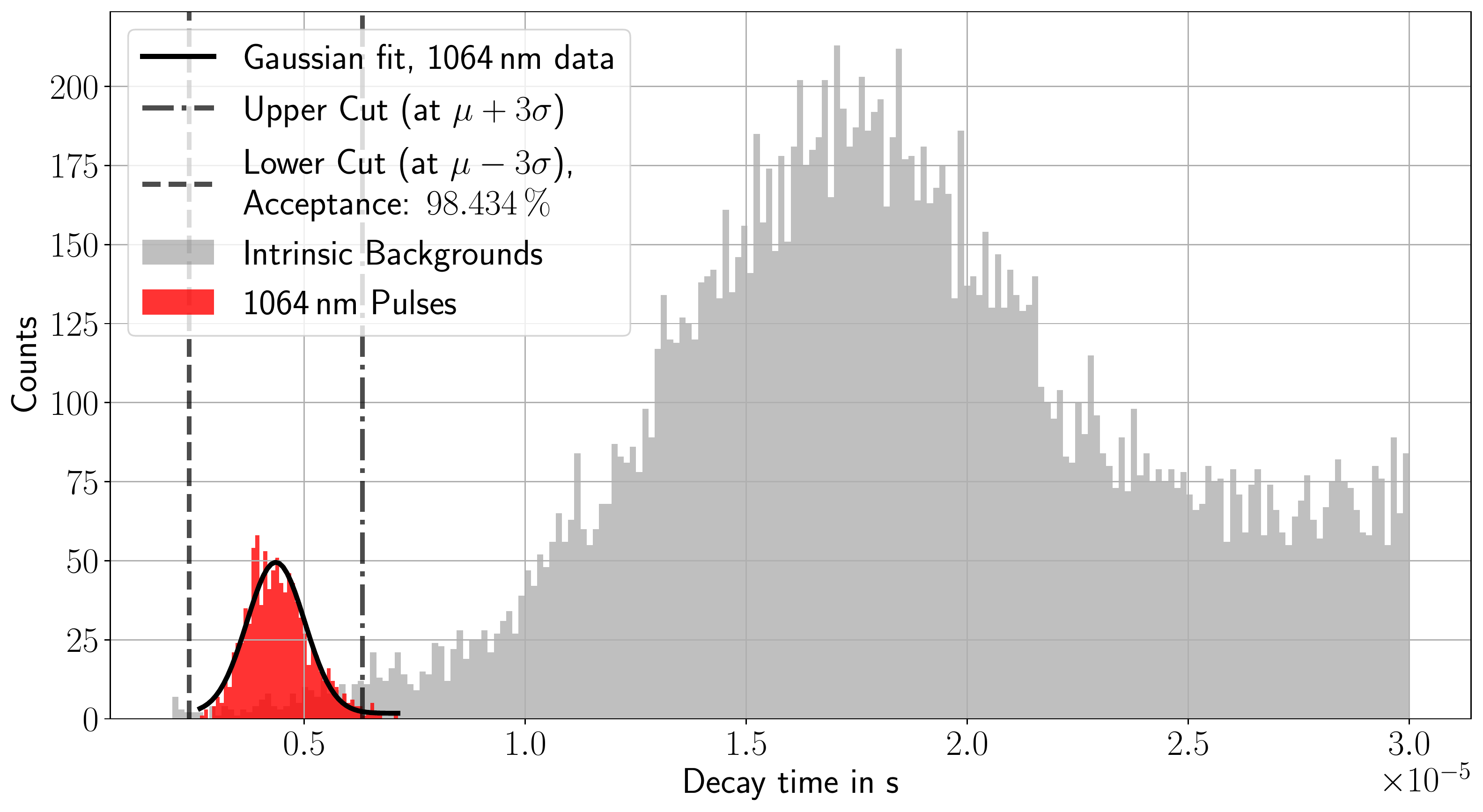}
			\caption{}
		\end{subfigure}
		\begin{subfigure}[t]{0.48\textwidth}
			\includegraphics[width=\textwidth]{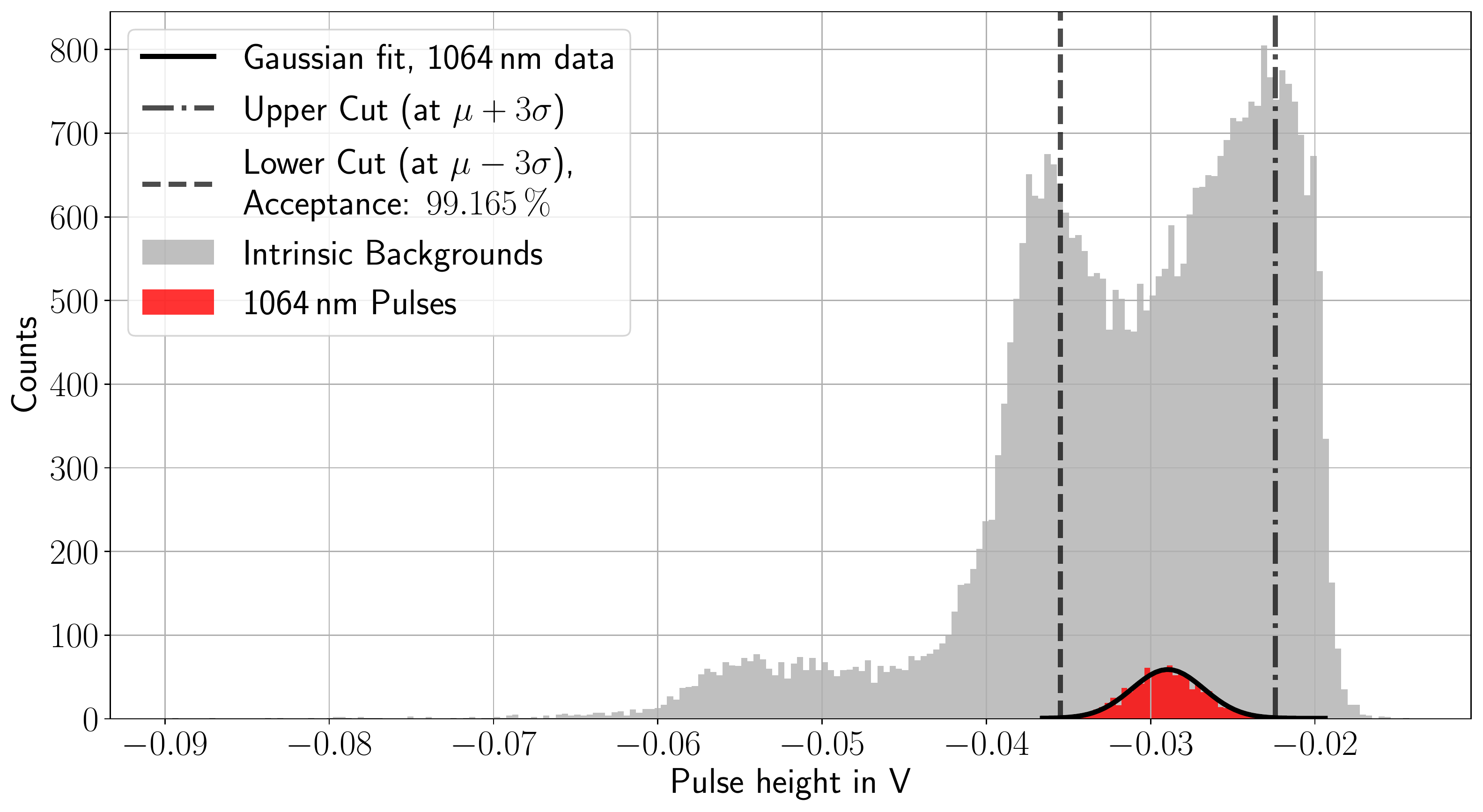}
			\caption{}
		\end{subfigure}
		\begin{subfigure}[t]{0.48\textwidth}
			\includegraphics[width=\textwidth]{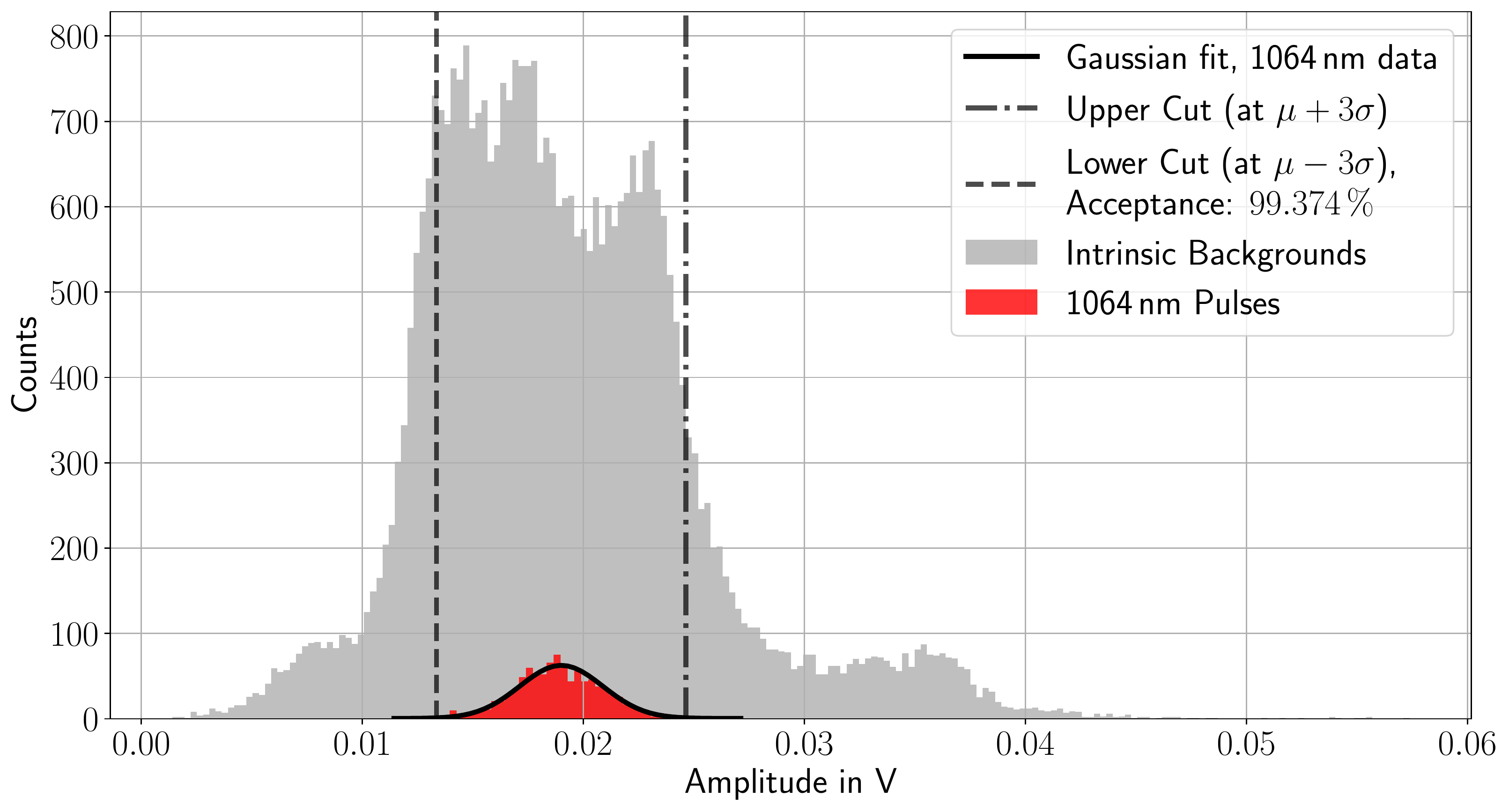}
			\caption{}
		\end{subfigure}
		\begin{subfigure}[t]{0.48\textwidth}
			\includegraphics[width=\textwidth]{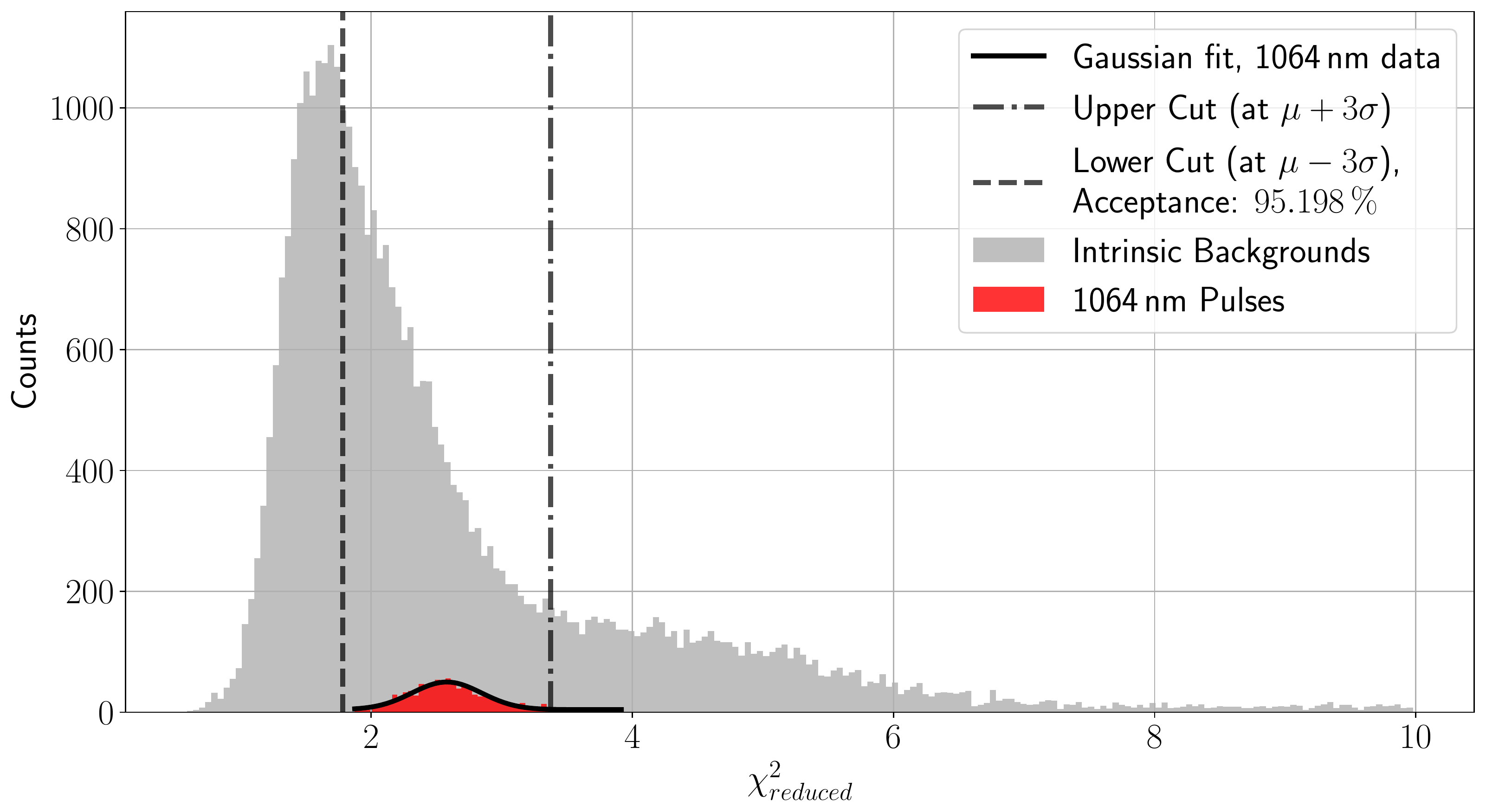}
			\caption{}
		\end{subfigure}
		
		\caption{The distributions for the pulse integral and other fit parameters of the intrinsic background events and 1064\,nm photons is shown, with exemplary cuts at $\mu \pm 3\sigma$. About 37,000 intrinsic background events (collected over 20 days), and $\sim$1,000 1064\,nm photon pulses (collected over $\sim$1\,s) are used here. The distribution of the pulse integral (theoretically proportional to the energy deposited in the TES) continues on below $-1 \cdot 10^{-6}$\,Vs but is not shown above. The background pulse integrals with magnitude lesser than the 1064\,nm photon pulses are typically dominated by electronic noise triggers, while those larger are dominated by pulses of other physical origins. }
		\label{Figure:TESPulseSelection}
	\end{figure}

	The fit parameters for 1064\,nm photons are used to establish a cut-region around the mean of the respective distributions in order to accept as many 1064\,nm photon pulses as possible while rejecting the background pulses.
	This selection region is contained by the bounds (or cuts) $\mu_{\text{fit param}} \pm n\cdot \sigma_{\text{fit param}}$, where $n$ is a positive number (Figure \ref{Figure:TESPulseSelection}).

	\begin{figure}[h]
		\centering
		\includegraphics[width=0.5\textwidth]{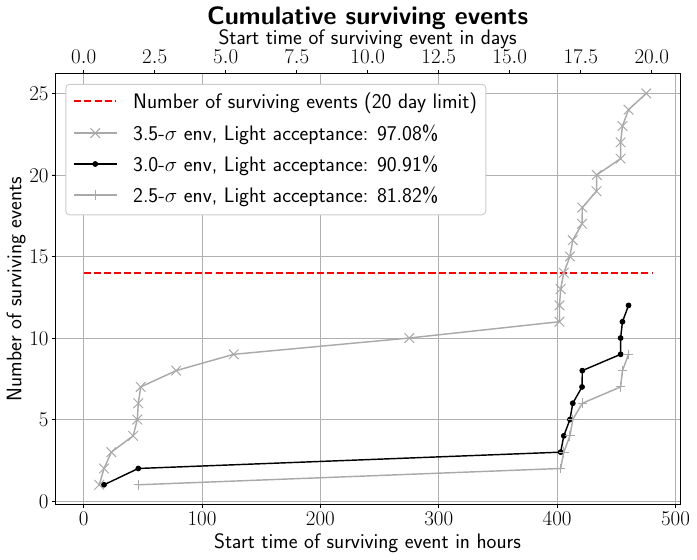}
		\caption{The number of intrinsic background pulses which survive the cuts $\mu_{\text{fit param}} \pm n\cdot \sigma_{\text{fit param}}$ over the 20-day data collection period, for various cuts, are shown. With the cut  used here we obtain 12 events over 20 days, lesser than the limit specified. The burst of photon-like events seen near the 400-hour mark is being investigated, they could originate from electromagnetic fields around the TES or some luminescence effects, and do not survive the cuts due to their difference from the goal photon regeneration rate of $\sim$1-2 photons per day.}
		\label{Figure:TESIntEvolution}
		
	\end{figure}
	For the TES to be viable for the ALPS II experiment, it must have a dark rate lesser than $7.7 \cdot 10^{-6}$\,cps over a 20-day period in order to detect 50 1064\,nm photons with a $5\sigma$ significance and 50\% detection efficiency \cite{ref_11}. Using our pulse selection approach for intrinsic background pulses collected over a period of 20 days, we obtain a dark rate of $6.9^{+2.62}_{-1.47} \cdot 10^{-6}$\,cps  with a choice of $n=3$ (Figure \ref{Figure:TESIntEvolution}).This entails a simultaneous 1064\,nm photon acceptance greater than 90\%.

	
	\section{Summary and Outlook}
	Considering only the intrinsic backgrounds, the TES is thus a viable candidate for use in the ALPS II experiment. The methodology used to establish cuts and select 1064\,nm-like pulses will be applied also to the extrinsic backgrounds, i.e. the background events triggered in a fiber-coupled system (compared to \cite{ref_12}, lower dark rates are already achieved). The efficiency of the fiber-to-TES coupling line is being measured using an independently calibrated power meter from the PTB in a specially designed setup, as in \cite{ref_13}, with promising results and further optimization steps to be undertaken. Approaches to reduce suspected sources of extrinsic backgrounds will also be investigated in the detector setup, viz. filtering, fiber-curling, etc. These novel techniques, as used in a cryogenic environment, are expected to significantly aid in  the reduction of background types that could otherwise lead to more background counts than can be tolerated. With a full simulation of the detector setup also being realized, the sources and influence of background events in the TES will be thoroughly investigated. The entire detector setup will be moved to the experimental site in HERA North, DESY, Hamburg, towards the end of 2022 where it will be characterised again before being implemented in the ALPS II experiment to detect axions and axion-like-particles, after the first science run using a heterodyne detection scheme \cite{ref_18}.
	
	\section{Acknowledgements}
	We want to thank NIST, USA, for the TES devices and PTB, Berlin, Germany, for the SQUID sensors, module and helpful advice. We would also like to thank Manuel Meyer and our ALPS collaborators.


\begin{thebibliography}{99}
		\bibitem{ref_1}
		Wilczek, Frank. "Problem of Strong P and T Invariance in the Presence of Instantons." Physical Review Letters 40.5 (1978): 279.
		\bibitem{ref_2}
		S. Weinberg, “A New Light Boson?”, Phys. Rev. Lett. 40, 223
		\bibitem{ref_3}
		R. D. Peccei, H. R. Quinn, “CP Conservation in the Presence of Pseudoparticles”, Phys. Rev. Lett. 38, 1440
		\bibitem{ref_4}
		Ringwald, Andreas. "Exploring the role of axions and other WISPs in the dark universe." Physics of the Dark Universe 1.1-2 (2012): 116-135.
		\bibitem{ref_5}
		Irastorza, Igor G., and Javier Redondo. "New experimental approaches in the search for axion-like particles." Progress in Particle and Nuclear Physics 102 (2018): 89-159.
		\bibitem{ref_6}
		P. Sikivie, “Experimental Tests of the "Invisible" Axion”, Phys. Rev. Lett. 51, 1415
		\bibitem{ref_7}
		Ehret, Klaus, et al. "New ALPS results on hidden-sector lightweights." Physics Letters B 689.4-5 (2010): 149-155.
		
		\bibitem{ref_8}
		Bähre, Robin, et al. "Any light particle search II—technical design report." Journal of Instrumentation 8.09 (2013): T09001.
		\bibitem{ref_9}
		A. E. Lita, B. Calkins, L. A. Pellouchoud, A. J. Miller, and S. Nam, “Superconducting transition-edge sensors optimized for high-efficiency photon number resolving detectors”, Society of Photo-Optical Instrumentation Engineers (SPIE) Conference Series, 7681, April 2010.
		\bibitem{ref_10}
		K. D. Irwin and G. C. Hilton, “Transition-edge sensors”, Cryogenic Particle Detection, pages 63–149, 2005
		\bibitem{ref_11}
		Jan Hendrik Põld and Hartmut Grote, "ALPS II-Design Requirement document", Document number v3, Internal Communication
		\bibitem{ref_12}
		Dreyling-Eschweiler, Jan, et al. "Characterization, 1064 nm photon signals and background events of a tungsten TES detector for the ALPS experiment." Journal of Modern Optics 62.14 (2015): 1132-1140.
		\bibitem{ref_13}
		M. Schmidt et al., “Photon-Number-Resolving Transition-Edge Sensors for the Metrology of Quantum Light Sources”, Journal of Low Temperature Physics volume 193, pages 1243–1250(2018)
		\bibitem{ref_14}
		Sokolov, Anton V., and Andreas Ringwald. "Photophilic hadronic axion from heavy magnetic monopoles." Journal of High Energy Physics 2021.6 (2021): 1-23.
		\bibitem{ref_17}
		Du, Peizhi, et al. "Sources of low-energy events in low-threshold dark matter detectors." arXiv preprint arXiv:2011.13939 (2020).
		\bibitem{ref_18}
		Bush, Zachary R., et al. "Coherent detection of ultraweak electromagnetic fields." Physical Review D 99.2 (2019): 022001.
		
	\end{thebibliography}
\end{document}